\documentclass[conference]{IEEEtran}
\IEEEoverridecommandlockouts
\usepackage{cite}
\usepackage{amsmath,amssymb,amsfonts}
\usepackage{algorithmic}
\usepackage{graphicx}
\usepackage{textcomp}
\usepackage{xcolor}
\def\BibTeX{{\rm B\kern-.05em{\sc i\kern-.025em b}\kern-.08em
    T\kern-.1667em\lower.7ex\hbox{E}\kern-.125emX}}
\begin{document}

\title{Pain Analysis, in Premature Infants, Using Near Infrared Spectroscopy (NIRS)
}

\author{\IEEEauthorblockN{1\textsuperscript{st} Matthew Compton}
\IEEEauthorblockA{\textit{ Division of Neonatology} \\
\textit{University of South Florida}\\
}
\and
\IEEEauthorblockN{2\textsuperscript{nd} Ghada Zamzmi}
\IEEEauthorblockA{\textit{Computer Science and Engineering} \\
\textit{University of South Florida}}
\and
\IEEEauthorblockN{3\textsuperscript{rd} Rahul Mhaskar}
\IEEEauthorblockA{\textit{Department of Internal Medicine} \\
\textit{University of South Florida}\\}
\and
\IEEEauthorblockN{4\textsuperscript{th} Maria Gieron}
\IEEEauthorblockA{\textit{Division of Pediatric Neurology} \\
\textit{University of South Florida}\\}
\and
\IEEEauthorblockN{5\textsuperscript{th} Marcia Kneusel}
\IEEEauthorblockA{\textit{Division of Neonatology} \\
\textit{University of South Florida}\\}
\and
\IEEEauthorblockN{5\textsuperscript{th} Judy Zarit}
\IEEEauthorblockA{\textit{Division of Neonatology} \\
\textit{University of South Florida}\\}
\and 
\IEEEauthorblockN{6\textsuperscript{th} Terri Ashmeade,}
\IEEEauthorblockA{\textit{Department of Pediatrics} \\
\textit{University of South Florida}\\}
}

\maketitle

\textbf{Background:} The role of neonatal pain on the developing nervous system is not completely understood, but
evidence suggests that sensory pathways are influenced by an infant’s pain experience.
Research has shown that an infant’s previous pain experiences lead to an increased, and likely
abnormal, response to subsequent painful stimuli. We are working to improve neonatal pain
detection through automated devices that continuously monitor an infant. The current study
outlines some of the initial steps we have taken to evaluate Near Infrared Spectroscopy (NIRS)
as a technology to detect neonatal pain. Our findings may provide neonatal intensive care unit (NICU) practitioners with the data necessary to monitor and perhaps better manage an
abnormal pain response.

\textbf{Methods:} A prospective pilot study was conducted to evaluate nociceptive evoked cortical activity in preterm infants. NIRS data were recorded for approximately 10 minutes prior to an acute painful
procedure and for approximately 10 minutes after the procedure. Individual data collection
events were performed at a weekly maximum frequency. Eligible infants included those
admitted to the Tampa General Hospital (TGH) NICU with a birth gestational age of less than 37
weeks.

\textbf{Results:} A total of 15 infants were enrolled and 25 individual studies were completed. Analysis demonstrated a statistically significant difference between the median of the pre- and post-painful
procedure data sets in each infant’s first NIRS collection (p value = 0.01).

\textbf{Conclusions:}
Initial analysis shows NIRS may be useful in detecting acute pain. An acute painful procedure is typically followed by a negative deflection in NIRS readings.

\section{Introduction}
The concept of neonatal pain has been quite fluid throughout history as perceptions and
evidence have slowly morphed into what we know and believe today. The study of neonatal
pain seems to have begun in the 1870s, when Dr. Flechsig proposed that it was unlikely that
neonates could experience pain because their neuronal myelination was not complete (Cope, 1998). Charles Darwin agreed with this view, when he wrote in his book The Expression of the Emotions in Man and Animals, that an infant’s pain expressions were related to reflexes only (Darwin, 1872). Even as late as the 1950s some pediatric surgeries were performed without
anesthesia and analgesia (Cope, 1998).

Today the prevailing belief is that not only do neonates experience pain, but their past pain experiences likely shape their futures responses to painful stimuli. This concept was first observed in infants who underwent circumcision and then were compared to age matched
controls receiving immunizations at four to six months of age. Researchers found that infants
who underwent circumcision demonstrated increased behavioral measures of pain compared to
their age-matched controls (Taddio, 1997). Likewise, infants of diabetic mothers who required
multiple heel lances in the first 24 hours of life demonstrated increased behavioral measures of
pain, compared to age-matched controls, when they received venipuncture for newborn screening samples at 24 hours of life (Taddio, 2002). 

Similar findings were demonstrated, in
evoked neuronal activity, when researchers found that premature infants yielded significantly
larger noxious evoked potentials compared to their age-matched term controls (Slater, 2010).
As the concept of neonatal pain has evolved, the tools used to measure it have changed.
Although behavioral measures remain the gold standard in the NICU, more objective and
continuous measurement methods are becoming available. One such technology is NIRS. NIRS
uses light in the 700 to 1000 nanometer range to penetrate skin, soft tissue, bone and brain to measure an infant’s deoxygenated and oxygenated hemoglobin concentrations (Ranger, 2011).
These measures have shown good correlation with behavioral measures of neonatal pain
(Slater, 2006).

In the current study we sought to detect neonatal pain using NIRS. This pilot study had the
ultimate goal of determining if NIRS is worthy of further investigation, and possible incorporation into a device, to continuously measure neonatal pain in the NICU.

\section{Methods}
We hypothesized that performing a painful procedure would result in a decrease in the regional
oximetry of the brain as measured by NIRS. The primary outcome measure was a change in
NIRS values, of the contralateral brain, following a painful procedure in the infant’s extremity.

\subsection{Infant Recruitment}
All consecutive premature infants, with a birth gestational age of less than 37 weeks, were
eligible for recruitment from the TGH NICU. The birth gestational age of each infant was obtained from obstetrical records, which included antenatal ultrasounds and maternal reports of last menstrual periods. Exclusion criteria included major congenital anomalies of the central nervous system or cardiovascular system, ongoing mechanical ventilation via an endotracheal tube and infants who had received analgesic or sedative drugs in the previous twenty four hours. This study was approved by the Tampa General Hospital Institutional Review Board. Figure 1 shows the enrolment numbers for this study.

\begin{figure}[t]
\includegraphics[width=1\linewidth]{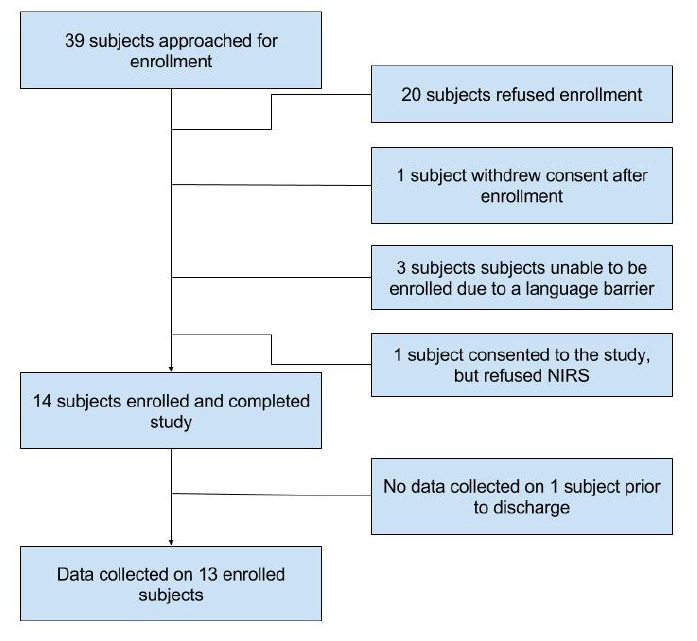}
\caption{Outcome of each infant considered for the current study.}
\label{fig}
\end{figure}

\subsection{Infant Data Collection}
Of the fifteen infants consented for the study, data were obtained on thirteen infants. Data were not collected on two infants as consent was withdrawn on one infant and one infant was
discharged prior to data collection. A total of twenty-five data collection events were performed on the fourteen infants that maintained informed consent. The number of data collection events per infant is displayed in Figure 2.

\begin{figure*}[t]
\includegraphics[width=1\linewidth]{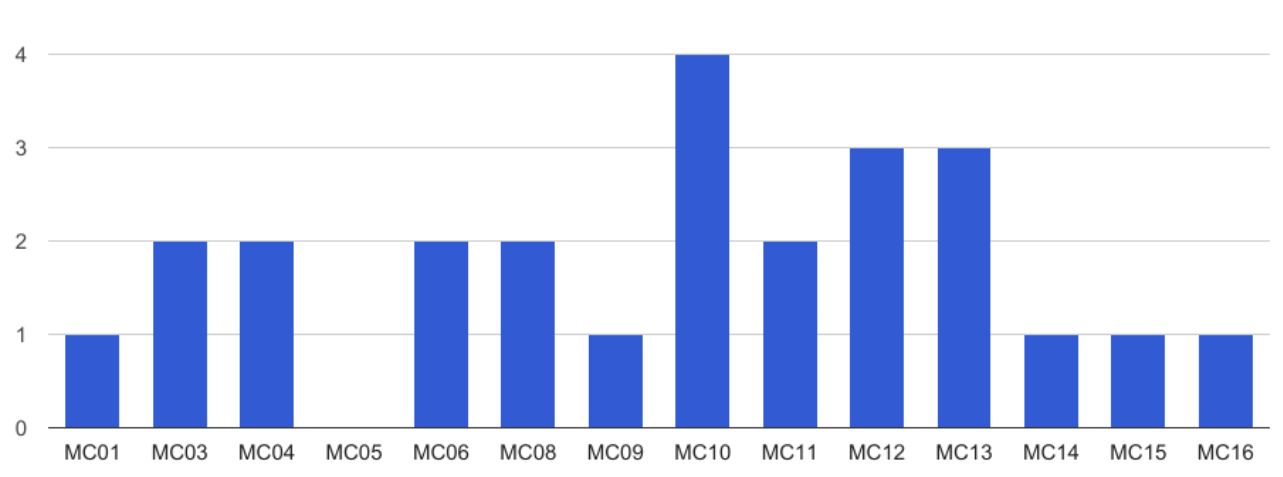}
\caption{The number of NIRS data collection events conducted on each enrolled infant.}
\label{fig}
\end{figure*}

\subsection{Study Equipment}
An INVOS 5100C NIRS meter was used to collect regional cerebral oximetry data. The meter
uses near infrared light to determine the deoxyhemoglobin (HbH) and the oxy-hemoglobin
(HbO2) concentrations. HbH and HbO2 are then added together to determine the total hemoglobin concentration (HbT). The meter then divides the HbO2 by the HbT to obtain a regional oximetry value (rSO2), which is displayed on the device’s screen. The meter’s sample
rate was 30 seconds. A VitalSync was used to timestamp and export the NIRS data into Excel. The VitalSync was also used to log events, such as the start of pre-procedure data collection and the start of the painful procedure.

\subsection{Study Protocol}
Once informed consent had been obtained, and no exclusion criteria were indentified, the
infant’s first data collection event was performed. Painful procedures included heal sticks for point-of-care glucose checks, routine labs, routine newborn metabolic screens, and
vaccinations. Additional data collection events were performed a minimum of one week apart.
No painful procedures were performed for the sole purpose of obtaining NIRS data. The study team reviewed the enrolled infants’ charts to determine the date and time of a painful
procedure and the bedside nurse performed it. 

In an effort to have the study’s results reflect usual practice, no alterations were made to the NICU’s usual method of obtaining labs or giving
vaccinations, and no alterations were made in the usual personnel performing the procedures.
Prior to the placing of the NIRS probe the team verified the location of the painful procedure with
the bedside nurse. The probe was then placed on the contralateral side of the forehead. Once the INVOS 5100C had established its auto-baseline, an event report was made in the
VitalSync and collection of the pre-procedure data was started. Pre-procedure data were collected for approximately ten minutes, at which point, the bedside nurse would start the painful procedure and an additional event report was made on the VitalSync. Post-procedure data were then collected for an additional 10 additional minutes after the painful procedure had ended.

\subsection{Statistical Analysis}
We investigated the change in the oximetry of the brain, as measured by NIRS, using Wilcoxson
signed rank test at a significance level of 0.05. We used SPSS version23 to conduct the
statistical analyses. In the analysis the ordered collections of each patient were compared to
each other (the 1st, 2nd, etc. collection events were compared), as we hypothesized that any
painful stimuli may impact the response to future painful stimuli.

\section{Results}
\subsection{Infant Demographics}

A total of thirty-nine infants were identified for enrolment. Of the thirty-nine identified infants consent was obtained for fifteen infants. Table 1 shows the patient demographics for the infants whose consent was not withdrawn. Two of the infants were found to have an intraventricular
hemorrhage (IVH).

\begin{figure}[t]
\includegraphics[width=1\linewidth]{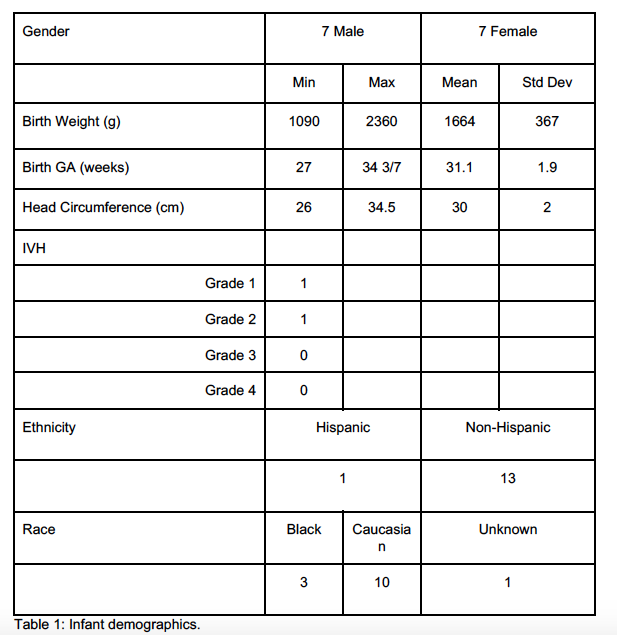}
\label{fig}
\end{figure}

\subsection{Data Analysis}
When comparing the first NIRS data collection event from each patient, we found a statistically
significant negative deflection in the regional cerebral oximetry values. The median of the medians had a pre-procedure value of 78.3\% and a post-procedure value of 76.5\% (p value =
0.01). Figure 3 plots the median pre- and post-procedure data values for each patient’s first data collection event, as well as, the median of the medians (displayed as Total). An insufficient
number of sequential data collection events hindered the determination of any meaningful conclusions about whether or not an infant’s NIRS response changes over time as the infant experiences more painful procedures.

\section{Discussion}
As previously described, several studies in infants have observed an increased response to painful stimuli over time. This is likely a learned response that results from a disruption in an infant’s ongoing nervous system development. We believe that continuous pain monitoring, with appropriate interventions, can alter this progression. This pilot study hypothesized that, following a painful procedure, a decrease in the contralateral regional cerebral oximetry would be seen. This hypothesis was based on literature findings that demonstrated that a painful stimulus results in an increase in the contralateral total hemoglobin concentration (Slater, 2006). 

Although the NIRS device used in our study computes a regional cerebral oximetry percentage, rather than displaying total hemoglobin concentration, our results
seem to agree with Slater et al. Since cortical activation, caused by a painful stimulus, likely
leads to a short relative decrease in the oxygenated hemoglobin concentration despite an increase in total blood flow, the end result is a decrease in the regional oximetry percentage. In
essence, the influx of oxygenated hemoglobin that should come with the increase in total
hemoglobin, is not initially sufficient to overcome the increased oxygen demands of the tissues.

While our results do seem to support our primary objective, this pilot study did not yield a
sufficient amount of data, due to low participant recruitment, to draw any meaningful conclusions about an infant’s cortical response to pain over time. In order to accomplish this
task, we propose that future studies should allow NIRS data to be continuously collected so that many acute procedures can be captured. Our study plan limited us to weekly measures, which led to less opportunities to collect data as time went on, as infants in the NICU tend to have most of their painful procedures performed early in their stays.

\begin{figure*}[t]
\includegraphics[width=1\linewidth]{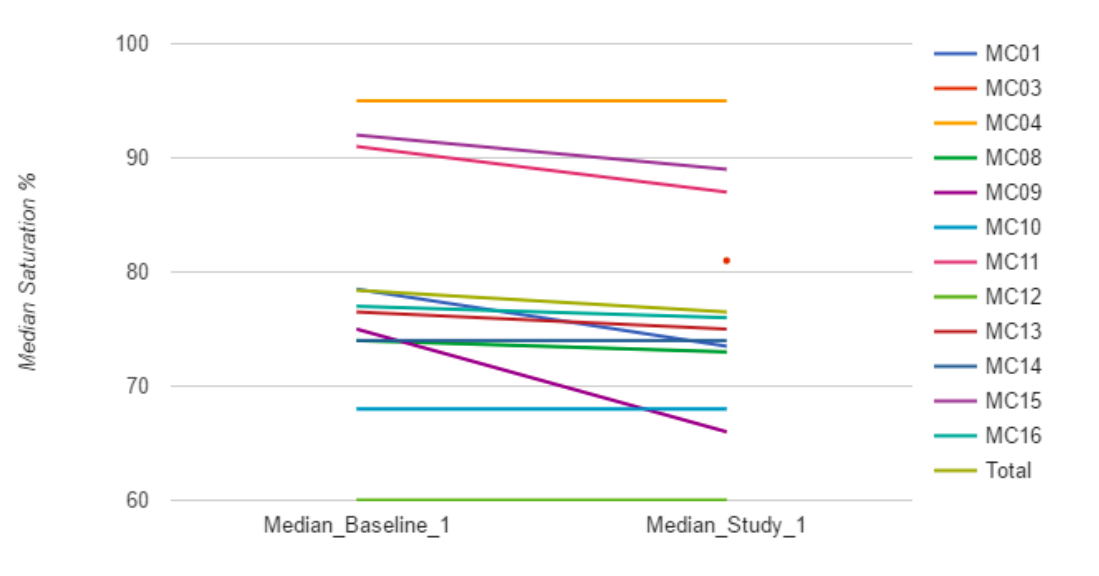}
\caption{This graph displays the pre and post-procedure median values for the first NIRS collection event for each infant. This
figure also includes the “Total,” which is the median of the medians for the 1st collection event.}
\label{fig}
\end{figure*}

The results of this pilot study are indeed promising as we believe they move us towards our ultimate goal of building a device that will continually monitor neonatal pain. We believe this information will allow NICU practitioners to better relieve neonatal pain thus breaking the
abnormal developmental cycle that currently ensues.

We have previously used facial recognition software and machine learning to detect behavioural measures of neonatal pain (Zamami). While the results of our previous study did reveal a high specificity for neonatal pain, the sensitivity was inadequate. It is our hope that incorporating NIRS will eventually result in improved sensitivity and therefore improved detection of neonatal
pain as we move forward.

\section*{References}
\begin{enumerate}
    \item Cope D (1998) Neonatal Pain: The Evolution of an Idea. American Society of
Anesthesiologists.

\item Ranger M, Johnston C, Limperopoulos C, Rennick J, du Plessis A (2011) Cerebral nearinfrared
spectroscopy as a measure of nociceptive evoked activity in critically ill infants.
Pain Research Management 16(5):331-336.

\item Slater R, Cantarella A, Gallella S, Worley A, Boyd S, Meek J, Fitzgerald M (2006)
Cortical pain responses in human infants. The Journal of Neuroscience 26(14):3662-
3666.

\item Slater R, Fabrizi L, Worley A, Meek J, Boyd S, Fitzgerald M (2010) Premature infants
display increased noxious-evoked neuronal activity in the brain compared to healthy
age-matched term-born infants. NeuroImage 583-589.

\item Slater R, Worley A, Fabrizi L, Roberts S, Meek J, Boyd S, Fitzgerald M (2010) Evoked
potentials generated by noxious stimulation in the human infant brain. European Journal
of Pain 14:321-326.

\item Slater R, Cantarella A, Franck L, Meek J, Fitzgerald M (2008) How well do clinical pain
assessment tools reflect pain in infants? PLOS Medicine 5(6):0928-0930.

\item Taddio A, Katz J, Ilersich A, Koren G (1997) Effect of neonatal circumcision on pain
response during subsequent routine vaccination. Lancet 349:599-603.

\item Taddio A, Shah V, Gilbert-MacLeod C, Katz J (2002) Conditioning and hyperalgesia in
newborns exposed to repeated heel lances. JAMA 288(7):857-861.

\item Zamzami, G, et al. “An approach for automated multimodal analysis of infants’ pain.”
Pattern Recognition (ICPR), 2016 23rd International Conference on. IEEE, 2016.

\item Darwin, C. (1872). The Expression of the Emotions in Man and Animals. (1st ed.).
London, UK: John Murray.
\end{enumerate}

\end{document}